\documentclass[english]{revtex4-1}
\usepackage[T1]{fontenc}
\usepackage[latin9]{inputenc}
\usepackage{mathtools}
\usepackage{amsmath}
\usepackage{graphicx}
\usepackage{wasysym}
\usepackage{esint}
\usepackage{babel}
\begin{document}
\title{Information Loss Pathways in a Numerically Exact Simulation of a non-Markovian
Open Quantum System }
\author{Evgeny A. Polyakov$^{1}$, Alexey N. Rubtsov$^{1,2}$}
\address{$^{1}$Russian Quantum Center, 100 Nonaya St., Skolkovo, Moscow 143025,
Russia}
\address{$^{2}$Department of Physics, Lomonosov Moskow State University, Leninskie
gory 1, 119991 Moscow, Russia}
\begin{abstract}
In the non-Markovian regime, the bath has the memory about the past
behaviour of the open quantum system. This memory has slowly-decaying
power-law tails. Such a long-range character of the memory complicates
the description of the resulting real-time dynamics on large time
scales. In a numerical simulation, this problem manifests itself in
a ``revival'', a spurious reflected signal which appears after a
finite time thus invalidating the simulation. In the present work,
we approach this problem and develop a numerical discretization of
the bath without revivals. We find that a crucial role is played by
the singularities of the bath spectral density (e.g. edges of bands):
the memory about the (spectral) behaviour of the open system in the
remote past is completely averaged out (forgotten), except an increasingly
small vicinity of these singular frequencies. Therefore, we introduce
the concept of memory channel, to denote such an irreversible information
loss proccess around a particular singular frequency. On a technical
side, we begin the treatment by noting that with respect to the memory
loss, the quantum field of the bath should be divided into the following
two different parts: the observable and the virtual quanta. Information
about the former is never lost: they contribute to the trace over
the bath degrees of freedom. The only way to avoid the revival from
the observable quanta is to calculate their dynamics exactly. We do
this by employing a stochastic sampling of the Husimi function of
the bath. The other part, the virtual quanta, are always annihilated
after a certain delay time. We construct a dedicated quantum representation
for the virtual states by assinging amplitudes to the delay times
before annihilation. It is in this representation we rigorously define
the concept of memory channels. 
\end{abstract}
\maketitle

\section{INTRODUCTION}

The description of a large number of physical phenomena is based on
the picture of a bounded quantum subsystem (the open quantum system,
OQS) which is coupled to a continuum (the bath). These include the
divesre physical scenarios like the atoms or quantum dots coupled
to a continuum of modes of phononic, electronic, and photonic reservoirs
\citep{Vega2017}; resonance phenomena in the electron-molecule collisions
\citep{Domcke1991,Muga2004,Tennyson2010,Zuev2014}, where the colliding
electron and the molecule form a joint interacting intermediary state
(the bounded subsystem), and the continuum is formed by outgoing (scattered)
states. In the reactive molecular collisions \citep{Neuhauser1989},
the intermediary state is coupled to the two continua: the incoming
states of reactants, and the outgoing states of products. In all the
cases the total problem has infinite dimensions, and due to the interactions
present in the subsystem, it has no trivial solution. The traditional
approach is to place a sharp boundary between the subsystem and the
continuum: the subsystem is interacting and nontrivial, but finite;
the continuum is infinite, but simple. Then, by (approximately) eliminating
the continual degress of freedom, one obtains a reduced finite description,
which is amenable to numerical solution. This is the essence of the
Feshbach projections method \citep{Zwanzig1960,Muga2004,Breuer2004,Breuer2007}. 

There is no problems with the Feshbach projections in the Markovian
mode. This regime is characterized by the two conditions: 1) the coupling
between the subsystem and the continuum is weak, and 2) the continuum
is much faster then the subsystem. The condition 2) is also may be
stated in a different way: that the background (continuum) spectral
density is flat on the scales of spectral features of the subsystem
\citep{Vega2017}. In operational terms, this often means that we
need to find such a boundary (between the subsystem and the continuum)
which fullfils these conditions \citep{Domcke1991}. In the case of
a success, the boundary between the continuum and the subsystem is
crossed instantly and only once by the incoming/outgoing particles
(bath excitations). This results in a simple reduced description.
For the open quantum systems, like atoms and quantum dots, the reduced
description amounts to the addition of a Lindbladian dissipator to
the von Neumann equation of the reduced density matrix \citep{Vega2017}.
In the case of resonant and reactive collisions, a local optical complex
potential \citep{Neuhauser1989,Domcke1991} is introduced into the
Hamiltonian. Thus, nowadays we have a fairly complete and consistent
picture of the Markovian mode. 

However, outside the validity of this mode, in the non-Markovian regime,
the concept of sharp boundary becomes problematic. The continuum acquires
a memory about the past subsystem behaviour. In other words, there
is non-zero amplitude that the particles (bath excitations) will return
back from the contuum to the subsystem: the emitted virtual photons
can be reabsorbed by the atom \citep{Vega2017}; the reaction fragments
can temporarily return to the interacting region during a reative
collision \citep{Estrada1989,Gertitschke1994,Kazansky1996,Plohn1998,Brems2002}.
Moreover, the memory of the continuum is long-range: no matter how
far the particles fly into the continuum, the return amplitude decays
only as inverse power law. This means that physically there is no
sharp boundary at all. And when we artificially place a sharp boundary,
the low energy spectrum of the problem will be distorted \citep{Domcke1991,Muga2004,Tennyson2010}.
In the time-resolved simulations, this corresponds to the revivals
after a finite time \citep{Hartman2017,Richter2017,Strathearn2018},
so that large-time asymptotical behaviour of the observables is corrupted.
When doing numerical computations, one may attempt to alleviate this
problem by shifting the boundary inside the continuum e. g. either
by a direct inclusion of the continuum modes or sites, or by shifting
the memory tail cutoff in the time-nonlocal approaches (quasi adiabatic
path integrals and related \citep{Makri1995,Richter2017,Strathearn2018}).
But then, the dimension of the reduced description becomes combinatorially
large, and we cannot advance much this way.

In summary, the long-range memory tails are the major obstacle for
the efficient description of low-energy and large-time physics of
open quantum systems. In the present paper we propose one possible
solution to this problem. 

Our treatment is based on our preceding paper \citep{Polyakov2018a}
where we have presented an alternative approach to the description
of open quantuam systems in a non-Markovian bath: we do not place
a sharp boundary between the retained and the eliminated degrees of
freedom in a reduced description. Instead, the quantum field of bath
excitations is divided into the following two components: 1) the excitations
which are irreversibly (only once) emitted by OQS, and which are observable
by a bath measurement, and 2) the virtual excitations which are always
reabsorbed by OQS before any measurement of the bath \citep{Polyakov2017b,Polyakov2017a}.
The former component, the observable quantum field, has a classical
stochastic structure, since its time evolution can be described by
a master equation for the probability distribution of the measurement
outcomes of the bath. Therefore, we can numerically simulate it by
Monte Carlo methods. The latter component, consisting of virtual excitations,
is a genuinely quantum object. Its time evolution should be computed
by a deterministic scheme. In other words, we obtain a reduced description
in which the virtual part of the bath state is retained, whereas the
observable part of the bath state is eliminated by addition of a classical
stochastic noise. One can understand this as a ``soft boundary''
between the OQS and the bath.

The two components of the bath quantum field differ from each other
in how they carry the memory about the past OQS behaviour. The observable
quanta directly contribute to the measurement, thus to the reduced
density matrix of OQS. Therefore, the information in the observable
quanta is not lost. In order to avoid the revivals from the observable
component, we propose to calculate its evolution exactly, by a Monte
Carlo simulation. At the same time, in the second component, the virtual
quanta, only the information in the return amplitudes is retained.
Therefore, there is a mechanism of the gradual loss of memory about
remote past. We identify this mechanism: it is based on a monotonically
increasing flattering of the memory tails \citep{Richter2017}. This
allows us to devise a ``soft'' discretization of the bath, so that
all the effects of the long-range tails are taken (numerically) into
account. 

Our approach is supported by the current developments in the literature
on simulation methods for the time-resolved simulations. In particular,
there are numerous combined stochastic-deterministic schemes \citep{Zhou2008,Yan2016}.
In these algorithms, the interaction of the bath with OQS is splitted
into the two parts: one is represented by a stochastic coloured noise
(whose correlator reproduces the bath memory function); the other
part is solved deterministically. These algorithms have demonstrated
promising convergence properties when simulating non-Markovian OQS,
at short to moderate time scales \citep{Zhou2008,Yan2016}. 

In section \ref{subsec:Model} we introduce the model of open quantum
system. We define the problem of long-range memory tails in section
\ref{subsec:Long-range-memory-tails}. Then, in section \ref{subsec:Two-types-of}
we briefly recapitulate our approach of observable and virtual quanta.
We demonstrate that the observable quanta can be stochastically simulated
without the revivals in section \ref{subsec:Problem-of-memory}. Then,
in section \ref{sec:AMPLITUDES-FOR-VIRTUAL} we develop a special
representation for the virtual states which helps us to identify the
memory loss mechanism in section \ref{subsec:Irreversibile-loss-of}.
In particular, we find that the memory loss is concentrated around
singularities of the bath spectral densities, which we call ``memory
channels''. This allows us to implement a ``soft'' discretization
of the bath, whose convergence properties are illustrated on a calculation
in section \ref{subsec:The-numerical-coarsegraining} We conclude
in section \ref{sec:CONCULSION}. In appendix \ref{sec:DERIVATION-OF-EQUATIONS}
we derive the equation of motions for OQS in the dedicated representation
for virtual states. In appendix \ref{sec:REGULARIZATION-OF-THE} we
provide details of how the channelwise memory functions are constructed.
Finally, in appendix \ref{sec:SPLIT-OPERATOR-SIMULATION} we provide
some detials about our implementation of the numerical methods

\section{MEMORY OF THE QUANTUM ENVIRONMENT}

We begin the exposition by introducing the model of open quantum system,
formulating the problem of long-range memory tails, and recapitulate
the approach of dressed quantum trajectories to it. 

\subsection{Model \label{subsec:Model}}

We consider a model of open system which is bilinearly coupled to
the bath of harmonic oscillators. The Hamiltonian is
\begin{equation}
\widehat{H}=\widehat{H}_{\textrm{s}}+\hat{s}\widehat{b}^{\dagger}+\hat{s}^{\dagger}\widehat{b}+\widehat{H}_{\textrm{b}},\label{eq:total_hamiltonian_schrodinger_picture}
\end{equation}
where $\widehat{H}_{\textrm{s}}$ is the OQS, $\widehat{H}_{\textrm{b}}$
is the bath
\begin{equation}
\widehat{H}_{\textrm{b}}=\intop_{0}^{+\infty}d\omega\omega\widehat{a}^{\dagger}\left(\omega\right)\widehat{a}\left(\omega\right).
\end{equation}
The coupling is through the operator $\widehat{s}$ which is in the
system's Hilbert space, and through the operator $\widehat{b}$ which
is in the Hilbert space of bath, 
\begin{equation}
\widehat{b}=\intop_{0}^{+\infty}d\omega c\left(\omega\right)\widehat{a}\left(\omega\right).
\end{equation}

\subsection{Long-range memory tails\label{subsec:Long-range-memory-tails}}

The memory function of the bath is defined in terms of the commutator
of the coupling operator (in the interaction picture)
\begin{equation}
M\left(\tau-\tau^{\prime}\right)=\left[\widehat{b}\left(\tau\right),\widehat{b}^{\dagger}\left(\tau^{\prime}\right)\right]=\intop_{0}^{+\infty}d\omega\left|c\left(\omega\right)\right|^{2}\exp\left(-i\omega\left(\tau-\tau^{\prime}\right)\right).\label{eq:bath_memory_function}
\end{equation}
For physical bath couplings (with no negative frequencies) the memory
function always has inverse power-law tails. This can be illustrated
by considering a typical form of the spectral coupling coefficient
\begin{equation}
c\left(\omega\right)=\sqrt{\frac{\alpha\omega_{\textrm{c}}}{2}}\left[\frac{\omega}{\omega_{\textrm{c}}}\right]^{\frac{s}{2}}\exp\left(-\frac{1}{2}\frac{\omega}{\omega_{\textrm{c}}}\right),\label{eq:typical_coupling}
\end{equation}
where $\alpha>0$ is the coupling strength; for $s<1$ we have the
so-called subohmic bath; $s=1$ corresponds to ohmic bath; finally,
the case $s>1$ is called the superohmic bath. The corresponding memory
function is 
\begin{equation}
M\left(\tau\right)=\frac{\alpha\omega_{c}^{2}}{2}\frac{\Gamma\left(s+1\right)}{\left(1+i\tau\omega_{c}\right)^{s+1}}\wasypropto\textrm{const}\times\tau^{-s-1}.\label{eq:typical_memory_function-1}
\end{equation}

\subsection{Two types of bath quanta\label{subsec:Two-types-of}}

Our goal is to devise a simulation scheme which takes into account
the full memory tails in a numerically exact way. Such a scheme would
enable us to perform long-time simulations without the revivals. Starting
from an initial factorized state
\begin{equation}
\left|\Psi\left(0\right)\right\rangle =\left|0\right\rangle _{\textrm{b}}\otimes\left|\psi\left(0\right)\right\rangle _{\textrm{s}},
\end{equation}
the open system begins to interact with the bath. This interaction
manifests itself in an emisstion and absorbtion of quanta (bath excitations).
Then, after a time $t$, we perform a non-selective measurement of
the bath state, which yields the reduced density matrix
\begin{equation}
\widehat{\rho}_{\textrm{s}}\left(t\right)=\textrm{Tr}_{\textrm{b}}\left|\Psi\left(t\right)\right\rangle \left\langle \Psi\left(t\right)\right|
\end{equation}
for the open system. Here the partial trace operation $\textrm{Tr}_{\textrm{b}}$
is done with respect to the bath degrees of freedom. From the point
of view of non-selective measurement (the $\textrm{Tr}_{\textrm{b}}$),
all the bath quanta emitted by OQS are divided into the following
two types: those which will survive up to the measurement time $t$
(the observable quanta), and those which will be absorbed by OQS (the
virtual quanta).

\subsection{Exact simulation of observable quanta}

The memory about the observable quanta is not lost, since these quanta
contribute to the $\textrm{Tr}_{\textrm{b}}$ and to the reduced density
matrix. There is no obvious way to take into account this ``observable''
memory on long times except a certain numerically exact simulation.
Any approximate simulation (e.g. finite discretization of the bath)
will yield strong revival signals. Due to infinite dimension of the
bath Hilbert space, a Monte Carlo simulation is the most appropriate
numerically exact approach. As was discussed in the preceding paper
\citep{Polyakov2018a}, the natural way to construct such a stochastic
simulation is to probabilistically sample the outcomes of a bath measurement.
By considering the measurement with respect to the coherent states
of the bath, in \citep{Polyakov2018a} the following Monte Carlo simulation
algorithm was presented. At a time moment $t$, the reduced density
matrix $\widehat{\rho}_{\textrm{s}}\left(t\right)$ of OQS is represented
as a statistical average of pure states,
\begin{equation}
\widehat{\rho}_{\textrm{s}}\left(t\right)=\overline{\left\{ \frac{\left\langle 0\right|_{\textrm{b}}\left|\Psi_{\textrm{dress}}\left(z,t\right)\right\rangle \left\langle \Psi_{\textrm{dress}}\left(z,t\right)\right|\left|0\right\rangle _{\textrm{b}}}{\left\Vert \left|\Psi_{\textrm{dress}}\left(z,0\right)\right\rangle \right\Vert ^{2}}\right\} }_{z},
\end{equation}
where there averaging is over the complex Gaussian spectral noise
$z\left(\omega\right)$, with the the statistics 
\begin{equation}
\overline{z\left(\omega\right)}=0,\,\,\,\overline{z\left(\omega\right)z\left(\omega^{\prime}\right)}=0,\,\,\,\overline{z\left(\omega\right)z^{*}\left(\omega^{\prime}\right)}=\delta\left(\omega-\omega^{\prime}\right).
\end{equation}
For each realization of the noize $z\left(\omega\right)$, the pure
state $\left|\Psi_{\textrm{dress}}\left(z,t\right)\right\rangle $
is a solution of the Schrodinger equation 
\begin{equation}
\left|\Psi_{\textrm{dress}}\left(z,t\right)\right\rangle =-i\widehat{H}_{\textrm{dress}}\left(z,t\right)\left|\Psi_{\textrm{dress}}\left(z,t\right)\right\rangle ,\label{eq:stochastic_schrodinger}
\end{equation}
where the noise-dependent Hamiltonian is 
\begin{equation}
\widehat{H}_{\textrm{dress}}\left(z,t\right)=\widehat{H}_{\textrm{s}}+\hat{s}\left(\xi\left(t\right)+\phi^{*}\left(t\right)+\widehat{b}^{\dagger}\left(t\right)\right)+\left(\hat{s}^{\dagger}-\overline{s}^{*}\left(t\right)\right)\widehat{b}\left(t\right).\label{eq:diffusion}
\end{equation}
Here, the time-dependent classical complex signal $\xi\left(t\right)$
depends on the noise,
\begin{equation}
\xi\left(t\right)=\intop_{0}^{+\infty}d\omega c^{*}\left(\omega\right)z\left(\omega\right)e^{i\omega t}.\label{eq:collapse_noise}
\end{equation}
Another complex time-dependent classical field 
\begin{equation}
\phi\left(t\right)=-i\intop_{0}^{t}d\tau M\left(t-\tau\right)\overline{s}\left(\tau\right)\label{eq:shift-2}
\end{equation}
self-consistently depends on the average of the coupling operator
$\widehat{s}$
\begin{equation}
\overline{s}\left(t\right)=\frac{\left\langle \Psi_{\textrm{dress}}\left(z,t\right)\left|0\right.\right\rangle _{\textrm{b}}\widehat{s}\left\langle 0\right|_{\textrm{b}}\left|\Psi_{\textrm{dress}}\left(z,t\right)\right\rangle }{\left\Vert \left\langle 0\right|_{\textrm{b}}\left|\Psi_{\textrm{dress}}\left(z,t\right)\right\rangle \right\Vert ^{2}}.\label{eq:source}
\end{equation}
at previous times. 

\subsection{Problem of memory tails for the virtual quanta\label{subsec:Problem-of-memory}}

Such an approach - the exact Monte-Carlo simulation of the observable
quantum fields - solves the revival problem for these observable fields.
In order to demonstrate this, in our previous paper \citep{Polyakov2018a}
we performed the numerical simulation of the driven two-level system
in a waveguide.

\begin{figure}
\includegraphics[scale=0.7]{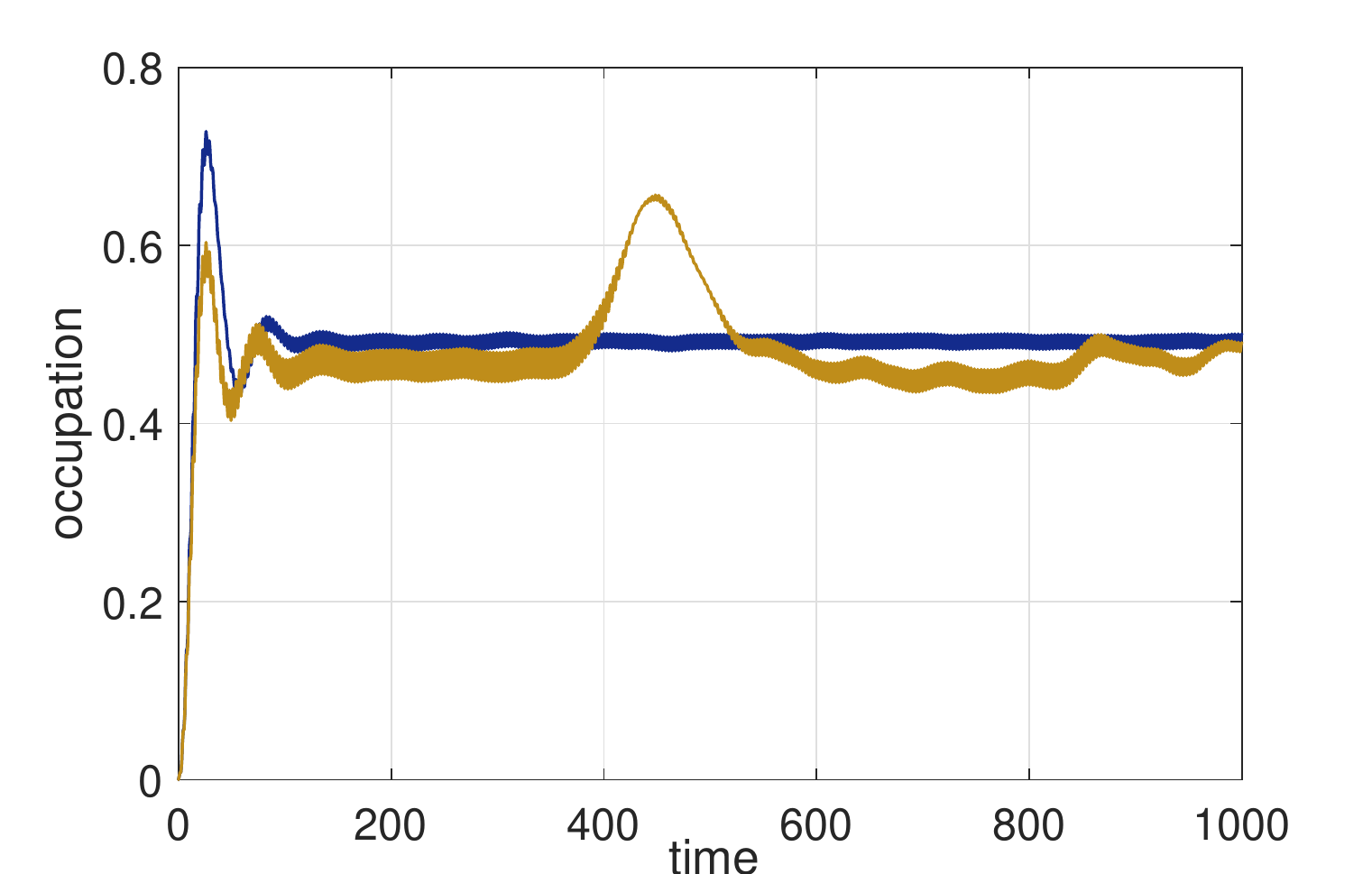}

\caption{\label{fig:zero_quanta_fixed_noise-1}In the previous paper \citep{Polyakov2018a},
a simulation of the driven two-level system in a waveguide was performed.
$n=0$ curve corresponds to the case of zero virtual quanta: no revivals
at extremely large times. $n=1$ curve corresponds to the case of
inclusion of one virtual quantum: a spurious revival signal is present
at times after $t=400$. }
\end{figure}
The Hamiltonian for the driven OQS is 

\begin{equation}
\widehat{H}_{\textrm{s}}=\varepsilon\widehat{\sigma}_{+}\widehat{\sigma}_{-}+\widehat{\sigma}_{+}f\left(t\right)+\widehat{\sigma}_{-}f^{*}\left(t\right),\label{eq:driven_spin_boson-1}
\end{equation}
with the driving field $f\left(t\right)=0.1\cos t$. The spin is coupled
to the waveguide through the spin lowering operator $\widehat{s}=\widehat{\sigma}_{-}$.
The degrees of freedom of the waveguide are the spectral modes labeled
by wavevector $k\in\left[0,\pi\right]$, with frequencies $\omega\left(k\right)=\varepsilon_{0}+2h\cos k$
and the coupling constant $c\left(k\right)=h\sqrt{\frac{2}{\pi}}\sin k$.
In Fig. \ref{fig:zero_quanta_fixed_noise-1} we present the simulation
results for $h=0.05$, $\varepsilon=1$ on large time scale. In this
figure, $n=0$ curve corresponds to the approximation when all the
virtual quanta are neglected (discarded all the operators $\widehat{b}\left(t\right)$,
$\widehat{b}^{\dagger}\left(t\right)$) in Eq. (\ref{eq:diffusion}).
We see that in this case, when there is only the observable quantum
field, the system reaches the steady state, and no revivals are present.

At the same time, if we impove this approximation and include one
virtual quantum in the simulation, which corresponds to $n=1$ curve
in Fig. \ref{fig:zero_quanta_fixed_noise-1}, and simulate the one-quantum
part of the dressed wavefuction $\Psi_{\textrm{dress}}\left(z,t\right)$
on a discretized grid of $N=20$ modes, we observe a clear revival
signal. See \citep{Polyakov2018a} for details.

We conclude that the stochastic simulation of observable field solves
only one half of the memory tail problem. The other half, for virtual
quanta, remains. In the next section we propose a solution to this
problem.

\section{\label{sec:AMPLITUDES-FOR-VIRTUAL}AMPLITUDES FOR VIRTUAL QUANTA}

In this section we deal with the question of how to efficiently compute
the projection
\begin{equation}
\left\langle 0\right|_{\textrm{b}}\left|\Psi_{\textrm{dress}}\left(z,t\right)\right\rangle \label{eq:vacuum_projection_conventional_approach}
\end{equation}
at long times. Usually, the state $\left|\Psi_{\textrm{dress}}\left(z,t\right)\right\rangle $
is represented as a vector of probability amplitudes over a certain
bath basis. The conventional choices for the basis are either the
frequency modes, or the sites of the equivalent semiinfinite chain
\citep{Prior2010,Woods2014,Vega2015}. However, the disadvantage of
this approach is that whenever we truncate the basis, a spurious reflected
signal (the revival) will come, thus invalidating the simulation after
a finite time. In order to delay this revival (in other words, to
increase the simulation time), the number of frequency modes (or sites)
should be increased linearly with time. This makes the long-time simulation
prohibitive, especially if a rather large number of virtual quanta
is retained. 

The solution to this problem begins with the following observation.
It is inefficient to perform the computation exactly according to
(\ref{eq:vacuum_projection_conventional_approach}). Because this
way we first solve for the full state $\left|\Psi_{\textrm{dress}}\left(z,t\right)\right\rangle $,
which contains both the observable and the virtual quanta. And only
when the observables are computed, we project it to $\left\langle 0_{\textrm{b}}\right|$,
thus discarding the observable quanta. Therefore, instead of $\left|\Psi_{\textrm{dress}}\left(z,t\right)\right\rangle $,
it is reasonable to find such a representation which contains \textit{only
}virtual quanta from the outset, and derive the equations of motion
directly in this representation.

\subsection{The delay-time amplitudes}

The characteristic property of the virtual quanta is that each of
them will be absorbed after a certian delay time. Then, assume we
have the bath in a superposition of number states 
\begin{equation}
\left|\varphi\left(t\right)\right\rangle =\varphi_{0}\left(t\right)\left|0\right\rangle _{\textrm{b}}+\left|\varphi_{1}\left(t\right)\right\rangle +\ldots+\left|\varphi_{N}\left(t\right)\right\rangle +\ldots
\end{equation}
For each number of quanta $N$, we introduce a set of delay times
$\tau_{1},\ldots\tau_{N}$, and define the delay-time amplitudes as
\begin{equation}
\varphi_{0}\left(t\right)=\left\langle 0\right|_{\textrm{b}}\left|\varphi\left(t\right)\right\rangle ,\label{eq:vacuum_delay_amplitude}
\end{equation}
\begin{equation}
\varphi_{1}\left(\tau_{1};t\right)=\left\langle 0\right|_{\textrm{b}}\widehat{b}\left(\tau_{1}\right)\left|\varphi\left(t\right)\right\rangle =\left\langle 0\right|_{\textrm{b}}\widehat{b}\left(\tau_{1}\right)\left|\varphi_{1}\left(t\right)\right\rangle ,\label{eq:one_quantum_delay_amplitude}
\end{equation}
\begin{equation}
\varphi_{2}\left(\tau_{1},\tau_{2};t\right)=\left\langle 0\right|_{\textrm{b}}\widehat{b}\left(\tau_{2}\right)\widehat{b}\left(\tau_{1}\right)\left|\varphi\left(t\right)\right\rangle =\left\langle 0\right|_{\textrm{b}}\widehat{b}\left(\tau_{2}\right)\widehat{b}\left(\tau_{1}\right)\left|\varphi_{2}\left(t\right)\right\rangle ,\label{eq:two_quantum_delay_amplitude}
\end{equation}
\[
\ldots
\]

\begin{equation}
\varphi_{N}\left(\tau_{1},\ldots,\tau_{N};t\right)=\left\langle 0\right|_{\textrm{b}}\widehat{b}\left(\tau_{N}\right)\ldots\widehat{b}\left(\tau_{1}\right)\left|\varphi\left(t\right)\right\rangle =\left\langle 0\right|_{\textrm{b}}\widehat{b}\left(\tau_{N}\right)\ldots\widehat{b}\left(\tau_{1}\right)\left|\varphi_{N}\left(t\right)\right\rangle .\label{eq:N_quantum_delay_amplitude}
\end{equation}
Here in $\varphi_{N}\left(\tau_{1},\ldots,\tau_{N};t\right)$ we use
the semicolon in order to distinguish the external (laboratory) time
$t$ from the dynamical variables - the delay times $\tau_{k}$. The
amplitude $\varphi_{N}\left(\tau_{1},\ldots,\tau_{N};t\right)$ is
a symmetric function of its arguments.

\subsection{Equations of motion in the delay-time picture}

Now let us write the equation for the dressed quantum trajectory (\ref{eq:stochastic_schrodinger})-(\ref{eq:source})
in terms of the delay-time amplitudes. In this case, besides the bath
degrees of freedom (now represented by delay times), we also have
the OQS quantum numbers. Denoting these additional quantum numbers
as $s$, we have the joint amplitude $\varphi_{N}\left(s,\tau_{1},\ldots\tau_{N};t\right)$
that all the virtual quanta will be absorbed after the delay times
$\tau_{1},\ldots,\tau_{N}$, \textit{and} OQS will be in the state
$\left|s\right\rangle $: 
\begin{equation}
\varphi_{N}\left(s,\tau_{1},\ldots\tau_{N};t\right)=\left\langle s\right|_{\textrm{s}}\otimes\left\langle 0\right|_{\textrm{b}}\widehat{b}\left(\tau_{N}\right)\ldots\widehat{b}\left(\tau_{1}\right)\left|\Psi_{\textrm{dress}}\left(z,t\right)\right\rangle .\label{eq:joint_delay_time_amplitude}
\end{equation}
We can also look on this from the other side, that we have a ket vector
$\left|\varphi_{N}\left(\tau_{1},\ldots\tau_{N};t\right)\right\rangle _{\textrm{s}}$
in the OQS Hilbert space, which depends on delay times, and which
is related to the joint amplitude as 
\begin{equation}
\left|\varphi_{N}\left(\tau_{1},\ldots\tau_{N};t\right)\right\rangle _{\textrm{s}}=\sum_{s}\varphi_{N}\left(s,\tau_{1},\ldots\tau_{N};t\right)\left|s\right\rangle _{\textrm{s}}.
\end{equation}
In order to derive the equations of motion for these delay-time ket
vectors, we differentiate with respect to time their definition (\ref{eq:joint_delay_time_amplitude}):
\begin{equation}
\partial_{t}\left|\varphi_{N}\left(\tau_{1},\ldots\tau_{N};t\right)\right\rangle _{\textrm{s}}=\left\langle 0\right|_{\textrm{b}}\widehat{b}\left(\tau_{N}\right)\ldots\widehat{b}\left(\tau_{1}\right)\left\{ -i\widehat{H}_{\textrm{dress}}\left(z,t\right)\right\} \left|\Psi_{\textrm{dress}}\left(z,t\right)\right\rangle .\label{eq:joint_delay_time_eom_start}
\end{equation}
Evaluating different terms in $\widehat{H}_{\textrm{dress}}$, we
obtain a hierarchy of equations, which we present here up to the second
order
\begin{equation}
\partial_{t}\left|\varphi_{\textrm{0}}\left(t\right)\right\rangle =-i\widehat{H}_{\textrm{stoch}}\left(t\right)\left|\varphi_{\textrm{0}}\left(t\right)\right\rangle -i\left(\widehat{s}^{\dagger}-\overline{s}^{*}\left(t\right)\right)\left|\varphi_{1}\left(0;t\right)\right\rangle ,\label{eq:delay_time_eq_1-1}
\end{equation}
\begin{multline}
\partial_{t}\left|\varphi_{\textrm{1}}\left(\tau;t\right)\right\rangle =-i\widehat{H}_{\textrm{stoch}}\left(t\right)\left|\varphi_{\textrm{1}}\left(\tau;t\right)\right\rangle +\partial_{\tau}\left|\varphi_{\textrm{1}}\left(\tau;t\right)\right\rangle \\
-iM\left(\tau\right)\widehat{s}\left|\varphi_{\textrm{0}}\left(t\right)\right\rangle -i\left(\widehat{s}^{\dagger}-\overline{s}^{*}\left(t\right)\right)\left|\varphi_{2}\left(0,\tau;t\right)+\varphi_{2}\left(\tau,0;t\right)\right\rangle ,
\end{multline}
\begin{multline}
\partial_{t}\left|\varphi_{\textrm{2}}\left(\tau_{1},\tau_{2};t\right)\right\rangle =-i\widehat{H}_{\textrm{stoch}}\left(t\right)\left|\varphi_{2}\left(\tau_{1},\tau_{2};t\right)\right\rangle +\left\{ \partial_{\tau_{1}}+\partial_{\tau_{2}}\right\} \left|\varphi_{2}\left(\tau_{1},\tau_{2};t\right)\right\rangle \\
-i\widehat{s}\frac{1}{2}\left\{ M\left(\tau_{1}\right)\left|\varphi_{\textrm{1}}\left(\tau_{2};t\right)\right\rangle +M\left(\tau_{2}\right)\left|\varphi_{\textrm{1}}\left(\tau_{1};t\right)\right\rangle \right\} ,\label{eq:delay_time_eq_3-2}
\end{multline}
where we have grouped into $\widehat{H}_{\textrm{stoch}}$ all the
terms which belong to the Hilbert space of OQS:
\begin{equation}
\widehat{H}_{\textrm{stoch}}\left(t\right)=\widehat{H}_{\textrm{s}}+\hat{s}\left(\xi\left(t\right)+\phi^{*}\left(t\right)\right).
\end{equation}
The average values of $\widehat{s}$ are now computed as 
\begin{equation}
\overline{s}\left(t\right)=\frac{\left\langle \varphi_{\textrm{0}}\left(t\right)\right|\widehat{s}\left|\varphi_{\textrm{0}}\left(t\right)\right\rangle }{\left\Vert \varphi_{\textrm{0}}\left(t\right)\right\Vert ^{2}}.
\end{equation}
We refer the interested reader to the appendix \ref{sec:DERIVATION-OF-EQUATIONS}
for the details of the derivation.

\subsection{Irreversibile loss of memory at large delay times\label{subsec:Irreversibile-loss-of}}

\begin{figure}
\includegraphics[scale=0.7]{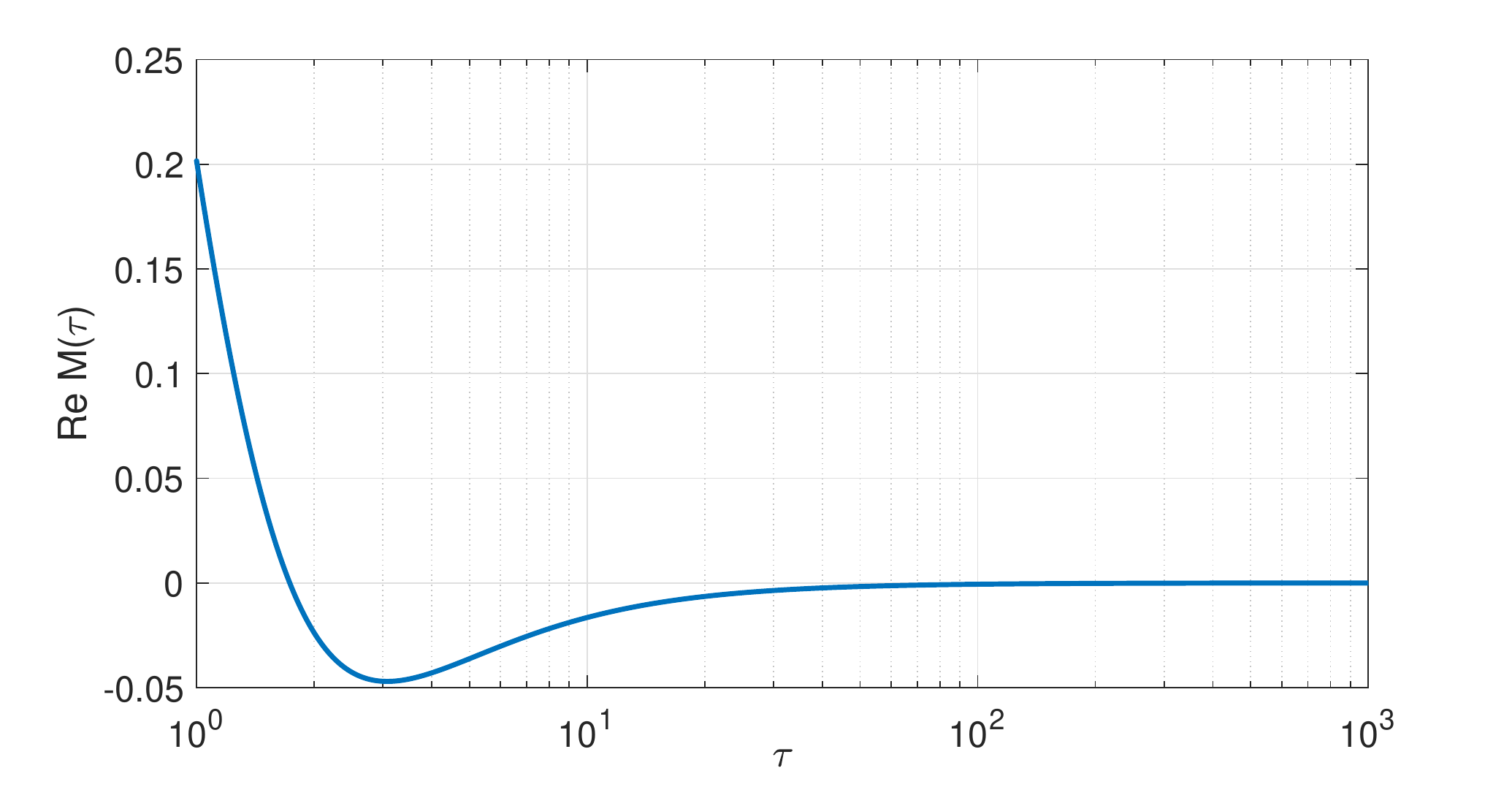}

\caption{\label{fig:Plot-of-the-}Plot of the real part of memory function
(\ref{eq:typical_memory_function-1})}

\end{figure}

Now let us analyze the equations (\ref{eq:delay_time_eq_1-1})-(\ref{eq:delay_time_eq_3-2}).
All the information which is not lost but which is accessible to the
outside world, is contained in the shifted classical noise $\xi\left(t\right)+\phi^{*}\left(t\right)$.
At the same time, the virtual quanta is a mechanism of the gradual
loss of information. In order to see how is it working, let us have
a look at a typical memory function, Eqs. (\ref{eq:typical_coupling})-(\ref{eq:typical_memory_function-1}).
In Fig. \ref{fig:Plot-of-the-} we plot real part of $M\left(\tau\right)$,
in logarithmic time scale. Imaginary part has qualitativly similar
behaviour. What we notice is that at large delay times the tails of
the memory function become increasingly flat. More formally, we can
say that with the increasing delay $\tau$, there is an increasingly
large time scale $\Delta\tau$ such that the memory function can be
considered effectively constant on the intervals $\left[\tau,\tau+\Delta\tau\right]$.
However, since in the equations (\ref{eq:delay_time_eq_1-1})-(\ref{eq:delay_time_eq_3-2})
the system states $\widehat{s}\left|\phi_{\textrm{0}}\left(t\right)\right\rangle $
and $\widehat{s}\left|\phi_{\textrm{1}}\left(\tau_{i};t\right)\right\rangle $
are continuously superimposed in the bath delay variables, with the
non-local weight $M\left(\tau\right)$, this means that at large delays
the information about OQS trajectory on the time scales smaller then
$\Delta\tau$, is averaged out (irreversibly lost). And asympotically,
at infinitely large delays, only the information about zero-frequency
behaviour of the system is retained. Therefore, the long-range irreversible
memory-loss mechanism is based on the monotonically increasing flattering
of the memory tails. 

\subsection{Case of a finite bandwidth}

\begin{figure}
\includegraphics[scale=0.7]{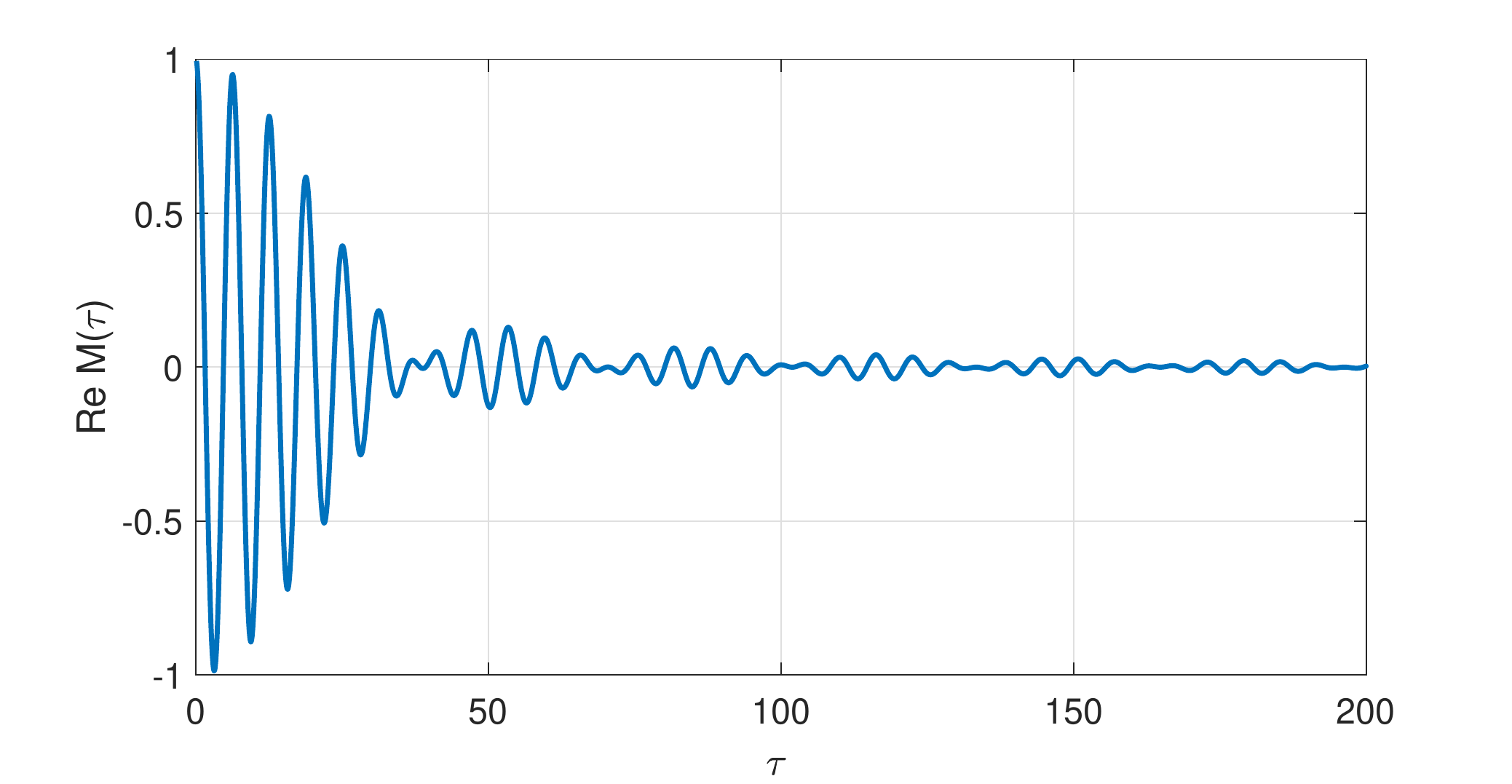}

\caption{\label{fig:band_limited_memory_tails}Real part of the memory function
of a semiinfinite waveguide. The waveguide is represented by a chain
of bose sites }

\end{figure}

Now let us return to the system treated in the section \ref{subsec:Problem-of-memory}.
The semiinfinite bath for that model has the memory function 
\begin{equation}
M\left(\tau\right)=e^{-i\varepsilon\tau}\frac{J_{1}\left(2h\tau\right)}{h\tau}.\label{eq:finite_band_memory_function}
\end{equation}
The behaviour of the tails in this case is a little bit harder. From
Fig. \ref{fig:band_limited_memory_tails} it is seen 
that while descreasing, the memory tails always oscillate, and we
cannot apply here directly the reasoning from the previous section:
there is no flattering of the tails. However, if we review other numerical
methods which employ the ideas of coarse-graining (e.g. the numerical
renormalization group method of Bulla et al. \citep{Bulla2008}),
we notice that in these methods all the memory (long-time behaviour
of the bath) is concentrated around the edges of the band, which is
reflected by the fact that we need exponentially-fine descrerization
of the bath spectrum at these points. And the oscillations in Fig.
\ref{fig:band_limited_memory_tails} are just the manifestation of
the interference between these two memory\textit{ channels}. In other
words, we expect that if we write the memory function separately for
each channel, then it will again have simple, increasingly-flattering
tails. 

Formally, we can define the notion of memory channels through the
large-delay asympotical behaviour of the memory function. In the case
of Eq. (\ref{eq:finite_band_memory_function}) we have
\begin{equation}
M\left(\tau\right)\sim e^{-i\left(\varepsilon-2h\right)t}\frac{e^{-i\pi/4}}{2i\sqrt{\pi}}\frac{1}{\left(ht\right)^{3/2}}+e^{-i\left(\varepsilon+2h\right)t}\frac{e^{+i\pi/4}}{2i\sqrt{\pi}}\frac{1}{\left(ht\right)^{3/2}}.
\end{equation}
Therefore, indeed, we have the two memory channels, associated to
the two momotonically-flattering power-law tails. Each channel is
attached to the frequency of the corresponding band edge,
\begin{equation}
\varepsilon_{1}=\varepsilon-2h,\,\,\,\textrm{and}\,\,\,\varepsilon_{2}=\varepsilon+2h.
\end{equation}
We expect that this feature is general: each edge of the band, and
each singularity in the spectral density, will lead to the existence
of its own channel. 

In order to make explicit this memory-channel structure, we decompose
the memory function: 
\begin{equation}
M\left(\tau\right)=e^{-i\varepsilon_{1}\tau}M_{1}\left(\tau\right)+e^{-i\varepsilon_{2}\tau}M_{2}\left(\tau\right),
\end{equation}
where 

\begin{equation}
M_{1}\left(\tau\right)=e^{-i2h\tau}\left(\frac{H_{1}^{\left(1\right)}\left(2h\tau\right)}{2h\tau}+iR\left(2h\tau\right)\right),\label{eq:channel_1}
\end{equation}
\begin{equation}
M_{2}\left(\tau\right)=e^{+i2h\tau}\left(\frac{H_{1}^{\left(2\right)}\left(2h\tau\right)}{2h\tau}-iR\left(2h\tau\right)\right),\label{eq:channel_2}
\end{equation}
and $R\left(z\right)$ is a certain fast-decaying real-valued function,
which fixes $\tau=0$ singularity of Hankel functions. See Appendix
\ref{sec:REGULARIZATION-OF-THE} for the details about $R\left(z\right)$. 

Let us rewrite the equations (\ref{eq:delay_time_eq_1-1})-(\ref{eq:delay_time_eq_3-2})
in terms of the channel memory functions. Since we now have the two
memory functions, the delay-time amplitudes get additional indices,
indicicating to which channel the delay argument referes to. It can
be interpreted as an additional ``quantum number'' of the virtual
quantum. So we obtain the set of wavefunctions: $\varphi_{\textrm{0}}\left(t\right)$;
$\phi_{\textrm{1}}^{\left(1\right)}\left(\tau;t\right)$ and $\phi_{\textrm{1}}^{\left(2\right)}\left(\tau;t\right)$;
$\phi_{2}^{\left(11\right)}\left(\tau_{1},\tau_{2};t\right)$, $\phi_{2}^{\left(12\right)}\left(\tau_{1},\tau_{2};t\right)$,
and $\phi_{2}^{\left(22\right)}\left(\tau_{1},\tau_{2};t\right)$;
.... For the vacuum of the bath we have
\begin{equation}
\partial_{t}\left|\varphi_{\textrm{0}}\left(t\right)\right\rangle =-i\widehat{H}_{\textrm{stoch}}\left(t\right)\left|\varphi_{\textrm{0}}\left(t\right)\right\rangle -i\left(\widehat{s}^{\dagger}-\overline{s}^{*}\left(t\right)\right)\sum_{k}e^{-i\varepsilon_{k}t}\left|\varphi_{1}^{\left(k\right)}\left(0;t\right)\right\rangle .\label{eq:delay_time_channel_1}
\end{equation}
For one virtual quantum in the $k$th channel we get
\begin{multline}
\partial_{t}\left|\phi_{\textrm{1}}^{\left(k\right)}\left(\tau;t\right)\right\rangle =-i\widehat{H}_{\textrm{stoch}}\left(t\right)\left|\phi_{\textrm{1}}^{\left(k\right)}\left(\tau;t\right)\right\rangle +\partial_{\tau}\left|\phi_{\textrm{1}}^{\left(k\right)}\left(\tau;t\right)\right\rangle -iM_{k}\left(\tau\right)e^{+i\varepsilon_{k}t}\widehat{s}\left|\phi_{\textrm{0}}\left(t\right)\right\rangle \\
-i\left(\widehat{s}^{\dagger}-\overline{s}^{*}\left(t\right)\right)e^{-i\varepsilon_{k}t}\left(\left|\phi_{2}^{\left(kk\right)}\left(0,\tau;t\right)\right\rangle +\left|\phi_{2}^{\left(kk\right)}\left(\tau,0;t\right)\right\rangle \right)-i\left(\widehat{s}^{\dagger}-s^{*}\left(t\right)\right)\sum_{l\neq k}e^{-i\varepsilon_{l}t}\left|\phi_{2}^{\left(kl\right)}\left(\tau,0;t\right)\right\rangle .\label{eq:delay_time_single_channel_quantum}
\end{multline}
Finally, the equations for the two quanta in the same channel $k$, 

\begin{multline}
\partial_{t}\left|\phi_{\textrm{2}}^{\left(kk\right)}\left(\tau_{1},\tau_{2};t\right)\right\rangle =-i\widehat{H}_{\textrm{stoch}}\left(t\right)\left|\phi_{2}^{\left(kk\right)}\left(\tau_{1},\tau_{2};t\right)\right\rangle +\left\{ \partial_{\tau_{1}}+\partial_{\tau_{2}}\right\} \left|\phi_{2}^{\left(kk\right)}\left(\tau_{1},\tau_{2};t\right)\right\rangle \\
-i\widehat{s}\frac{e^{+i\varepsilon_{k}t}}{2}\left\{ M_{k}\left(\tau_{1}\right)\left|\phi_{\textrm{1}}\left(\tau_{2};t\right)\right\rangle +M_{k}\left(\tau_{2}\right)\left|\phi_{\textrm{1}}\left(\tau_{1};t\right)\right\rangle \right\} ,\label{eq:delay_time_eq_3-1}
\end{multline}
and for the two quanta in different channels $k\neq l$ 
\begin{multline}
\partial_{t}\left|\phi_{\textrm{2}}^{\left(kl\right)}\left(\tau_{1},\tau_{2};t\right)\right\rangle =-i\widehat{H}_{\textrm{stoch}}\left(t\right)\left|\phi_{2}^{\left(kl\right)}\left(\tau_{1},\tau_{2};t\right)\right\rangle +\left\{ \partial_{\tau_{1}}+\partial_{\tau_{2}}\right\} \left|\phi_{2}^{\left(kl\right)}\left(\tau_{1},\tau_{2};t\right)\right\rangle \\
-i\widehat{s}\left\{ e^{+i\varepsilon_{k}t}M_{k}\left(\tau_{1}\right)\left|\phi_{\textrm{1}}^{\left(l\right)}\left(\tau_{2};t\right)\right\rangle +e^{+i\varepsilon_{l}t}M_{l}\left(\tau_{2}\right)\left|\phi_{\textrm{1}}^{\left(k\right)}\left(\tau_{1};t\right)\right\rangle \right\} .\label{eq:delay_time_channel_4}
\end{multline}

\subsection{The numerical coarsegraining of delay-time amplitudes\label{subsec:The-numerical-coarsegraining}}

Now its time to implement our understanding of the irreversible memory
loss mechanism. Since the time-scales of the memory functions increase
at large delay-times, we ``compress'' them by making a substitution
\begin{equation}
\tau=\tau\left(\varsigma\right)\coloneqq\frac{1}{2h}\left(as+b\left\{ e^{\left(\frac{\varsigma}{c}\right)^{2}}-1\right\} \right),\label{eq:cg_substitution}
\end{equation}
where in our numerical calculations we employ $a=0.5$, $b=0.1$,
$c=0.9$. Actually, we didn't try to find the most optimal substitution.
The form (\ref{eq:cg_substitution}) was guessed by eye: so that the
memory function is not too compressed near $\tau=0$, and at the same
time, that the power-law tails are sufficiently pursed. Then, the
memory function is considered as a function of the independent argument
$\varsigma$:
\begin{equation}
M_{k}\left(\varsigma\right)\coloneqq M_{k}\left(z\left(\varsigma\right)\right).
\end{equation}
To express the equations (\ref{eq:delay_time_channel_1})-(\ref{eq:delay_time_channel_4})
in terms of the $\varsigma$-variables, we make the following substitutions:
\begin{equation}
M_{k}\left(\tau\right)\to M_{k}\left(\varsigma\right),\,\,\,\phi_{\textrm{1}}^{\left(k\right)}\left(\tau;t\right)\to\phi_{\textrm{1}}^{\left(k\right)}\left(\varsigma;t\right),\,\,\,\phi_{2}^{\left(kl\right)}\left(\tau_{1},\tau_{2};t\right)\to\phi_{2}^{\left(kl\right)}\left(\varsigma_{1},\varsigma_{2};t\right),
\end{equation}
and 
\begin{equation}
\partial_{\tau}\to\left[\partial_{\varsigma}\tau\left(\varsigma\right)\right]^{-1}\partial_{\varsigma}.
\end{equation}
A grid of $\varsigma$ values was employed: 
\begin{equation}
\varsigma_{1}=0,..\varsigma_{2}=\Delta\varsigma,..\varsigma_{m}=\varsigma_{\textrm{max}},
\end{equation}
where
\begin{equation}
\varsigma_{\textrm{max}}=1,\,\,\,\Delta\varsigma=\varsigma_{\textrm{max}}/\left(m-1\right).
\end{equation}
and
\begin{equation}
m=7.
\end{equation}
We have employed the cutoff value
\begin{equation}
\varsigma_{\textrm{max}}=2.8525.
\end{equation}
All the amplitudes depending on delay times were approximated as piecewise
polynomials on the intervals $\left[\varsigma_{i},\varsigma_{i+1}\right]$.
The order and the coefficients of the polynomials where determined
from the requirement to reproduce the values and first three derivatives
of the amplitudes on the grid $\varsigma_{i}$. Yet, we don not argue
that this is the most efficient coarse-graining and interpolation
scheme. Rather, in this work we provide it as a proof-of-principle. 

\begin{figure}
\includegraphics[scale=0.7]{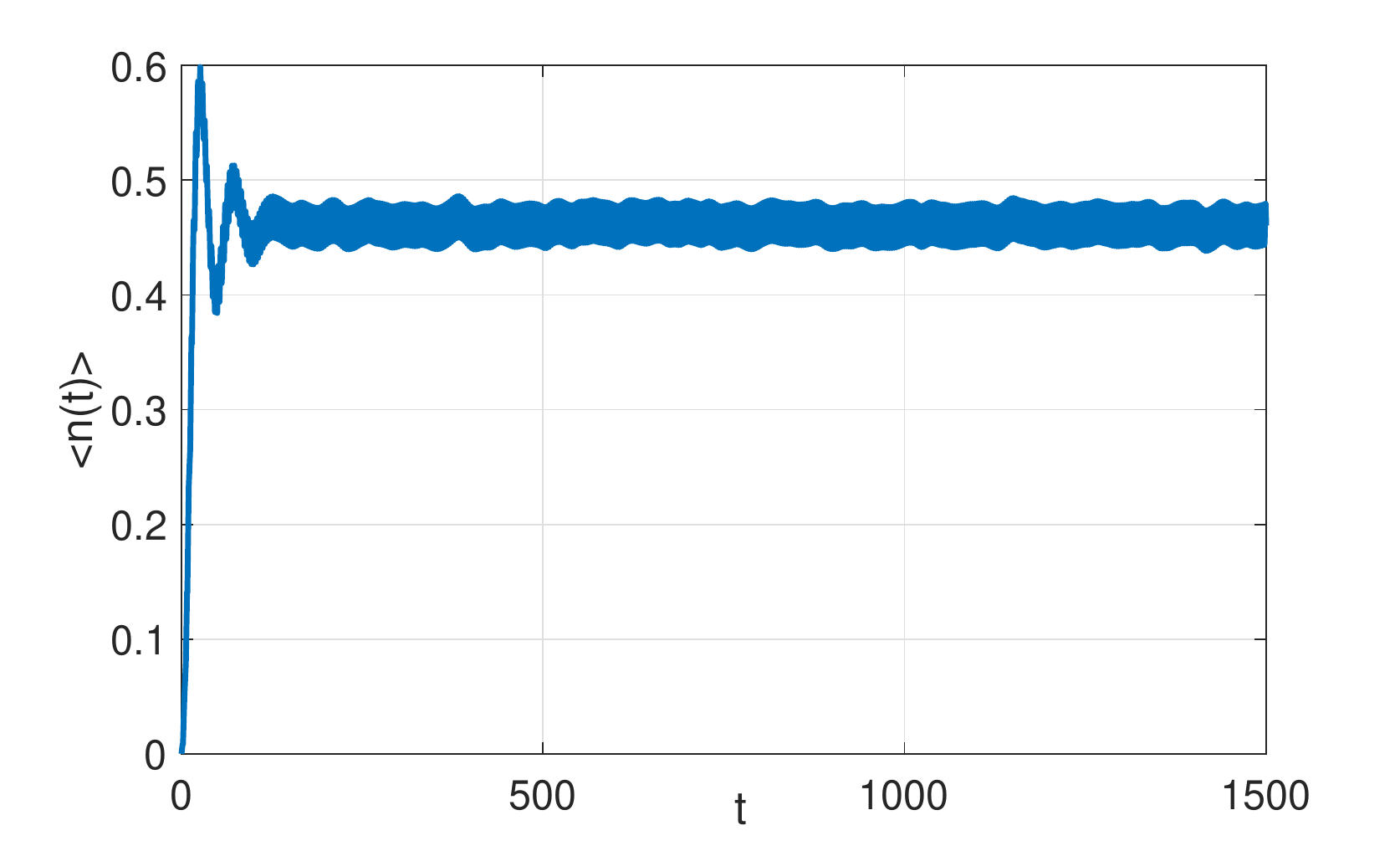}

\caption{\label{fig:Simulation-of-the}Simulation of the driven two-level system
in a wave-guide, described in section \ref{subsec:Problem-of-memory}.
Two virtual quanta are taken into account. When a soft coarse-graining
of the memory function on large delay times is performed, the revivals
dissappear. There is no deterioration of accuracy on large time scales. }

\end{figure}

The resulting system of equations was solved by a second-order symmetric
split-operator approach, where we considered separately the propagations
generated by free shift in delay time, by hopping with the memory
function $M_{k}\left(\tau\right)$, by absorbtion of virtual quantum,
and by the Schrodinger evolution. See the Appendix \ref{sec:SPLIT-OPERATOR-SIMULATION}
for the details about the slip-operator method. The results are presented
in Fig. \ref{fig:Simulation-of-the}
From this graph we see that the system successfully reaches the stationary
non-equilibrium state. There is no revivals or other deterioration
of the accuracy of observables. 

\section{\label{sec:CONCULSION}CONCULSION }

In this work we present a solution for the problem of how to discretize
the bath so that there is no revivals at large times. We have found
a soft coarse graining procedure for the memory function so that the
whole effect of its long-range tails is taken into account, in a numerically
exact way. We illustrate this approach by providing the results of
test calculations for the driven spin-boson model: only two virtual
excitations are enough to achieve the uniform physical convergence
on a large time interval.

Another interesting result is that each singularity in the spectral
density of the environment (e.g. band boundary) leads to the existence
of its own memory channel, which ``operates'' on the frequency of
the singularity. The physical content of this result is that at large
delay times, the memory about the past behaviour of OQS is retained
only in the vanishingly small vicinity of these singular frequencies,
and is completely averaged out (forgotten) away from this discrete
set of spectral lines. 

Now the actual task is to extend this result to a strong coupling
regime, where a large number of virtual quanta may be present. One
of the prossible approaches is to implement a singular value decomposition
of the coarse-grained delay-time wavefunction of virtual quanta, in
order to further ``compress'' the virtual quanta wavefunction. This
is a matter of future investigation.
\begin{acknowledgments}
The study was founded by the RSF, grant 16-42-01057.
\end{acknowledgments}

\appendix

\section{\label{sec:DERIVATION-OF-EQUATIONS}DERIVATION OF EQUATIONS IN THE
DELAY-TIME PICTURE}

In order to derive the equations of motion for the delay time amplitudes,
we need to consider the following three ingredients of $\widehat{H}_{\textrm{dress}}\left(z,t\right)$:
the free bath motion, the action of the annihilation operator, and
the action of the creation operator.

Let us derive the equation of motion for the delay-time amplitudes
under a free evolution of the bath:
\begin{multline}
\partial_{t}\varphi_{N}\left(\tau_{1},\ldots\tau_{p}\ldots,\tau_{N};t\right)=-i\left\langle 0\right|_{\textrm{b}}\widehat{b}\left(\tau_{N}\right)\ldots\widehat{b}\left(\tau_{p}\right)\ldots\widehat{b}\left(\tau_{1}\right)\widehat{H}_{\textrm{b}}\left|\varphi_{N}\left(t\right)\right\rangle \\
=-i\left\langle 0\right|_{\textrm{b}}\widehat{b}\left(\tau_{N}\right)\ldots\widehat{b}\left(\tau_{p}\right)\ldots\left[\widehat{b}\left(\tau_{1}\right),\widehat{H}_{\textrm{b}}\right]\left|\varphi_{N}\left(t\right)\right\rangle -\ldots-i\left\langle 0\right|_{\textrm{b}}\widehat{b}\left(\tau_{N}\right)\ldots\left[\widehat{b}\left(\tau_{p}\right),\widehat{H}_{\textrm{b}}\right]\ldots\widehat{b}\left(\tau_{1}\right)\left|\varphi_{N}\left(t\right)\right\rangle \\
\ldots-i\left\langle 0\right|_{\textrm{b}}\left[\widehat{b}\left(\tau_{N}\right),\widehat{H}_{\textrm{b}}\right]\ldots\widehat{b}\left(\tau_{p}\right)\ldots\ldots\widehat{b}\left(\tau_{1}\right)\left|\varphi_{N}\left(t\right)\right\rangle \\
=\sum_{k=1}^{N}\partial_{\tau_{k}}\varphi_{N}\left(\tau_{1},\ldots,\tau_{N};t\right).\label{eq:free_evolution}
\end{multline}
Here, in the first line we have employed the Schrodinger equation
for the free bath state $\left|\varphi_{N}\left(t\right)\right\rangle $.
Then, we moved $\widehat{H}_{\textrm{b}}$ to the left by commuting
it through all $\widehat{b}\left(\tau_{p}\right)$. Finally, in order
to arrive at the last line, we use the properties $\left\langle 0\right|_{\textrm{b}}\widehat{H}_{\textrm{b}}=0$
and $\left[\widehat{b}\left(\tau\right),\widehat{H}_{\textrm{b}}\right]=i\partial_{\tau}\widehat{b}\left(\tau\right).$

To describe the action of the annihilation operator $\widehat{b}$,
we introduce a ``quantum-subtracted'' bath state 
\begin{equation}
\left|\varphi^{\left(-1\right)}\left(t\right)\right\rangle =\widehat{b}\left|\varphi\left(t\right)\right\rangle .
\end{equation}
Since $\widehat{b}=\widehat{b}\left(0\right)$, their delay-time amplitudes
are related as
\begin{equation}
\varphi_{0}^{\left(-1\right)}\left(t\right)=\left\langle 0\right|_{\textrm{b}}\left|\varphi^{\left(-1\right)}\left(t\right)\right\rangle =\left\langle 0\right|_{\textrm{b}}\widehat{b}\left(0\right)\left|\varphi\left(t\right)\right\rangle =\varphi_{1}\left(0;t\right),
\end{equation}
\begin{equation}
\varphi_{1}^{\left(-1\right)}\left(\tau_{1};t\right)=\left\langle 0\right|_{\textrm{b}}\widehat{b}\left(\tau_{1}\right)\left|\varphi^{\left(-1\right)}\left(t\right)\right\rangle =\left\langle 0\right|_{\textrm{b}}\widehat{b}\left(\tau_{1}\right)\widehat{b}\left(0\right)\left|\varphi\left(t\right)\right\rangle =\varphi_{2}\left(\tau_{1},0;t\right),
\end{equation}
\[
\ldots
\]
\begin{multline}
\varphi_{N}^{\left(-1\right)}\left(\tau_{1},\ldots,\tau_{N};t\right)=\left\langle 0\right|_{\textrm{b}}\widehat{b}\left(\tau_{N}\right)\ldots\widehat{b}\left(\tau_{1}\right)\left|\varphi^{\left(-1\right)}\left(t\right)\right\rangle \\
=\left\langle 0\right|_{\textrm{b}}\widehat{b}\left(\tau_{N}\right)\ldots\widehat{b}\left(\tau_{1}\right)\widehat{b}\left(0\right)\left|\varphi\left(t\right)\right\rangle =\varphi_{N+1}\left(\tau_{1},\ldots,\tau_{N},0;t\right).
\end{multline}
In order to derive the representation of the last element, the action
of the creation operator, we introduce a ``quantum-added'' bath
state 
\begin{equation}
\left|\varphi^{\left(+1\right)}\left(t\right)\right\rangle =\widehat{b}^{\dagger}\left|\varphi\left(t\right)\right\rangle ,
\end{equation}
 and employ the two-time canonical commutation relation (\ref{eq:bath_memory_function}):
\begin{equation}
\varphi_{0}^{\left(+1\right)}\left(t\right)=\left\langle 0\right|_{\textrm{b}}\left|\varphi^{\left(+1\right)}\left(t\right)\right\rangle =\left\langle 0\right|_{\textrm{b}}\widehat{b}^{\dagger}\left|\varphi\left(t\right)\right\rangle =0,
\end{equation}
\begin{multline}
\varphi_{1}^{\left(+1\right)}\left(\tau_{1};t\right)=\left\langle 0\right|_{\textrm{b}}\widehat{b}\left(\tau_{1}\right)\left|\varphi^{\left(+1\right)}\left(t\right)\right\rangle =\left\langle 0\right|_{\textrm{b}}\widehat{b}\left(\tau_{1}\right)\widehat{b}^{\dagger}\left|\varphi\left(t\right)\right\rangle =M\left(\tau_{1}\right)\left\langle 0\right|_{\textrm{b}}\left|\varphi\left(t\right)\right\rangle =M\left(\tau_{1}\right)\varphi_{0}\left(t\right),
\end{multline}
\[
\ldots
\]
\begin{multline}
\varphi_{N}^{\left(+1\right)}\left(\tau_{1},\ldots,\tau_{N};t\right)=\left\langle 0\right|_{\textrm{b}}\widehat{b}\left(\tau_{N}\right)\ldots\widehat{b}\left(\tau_{1}\right)\left|\varphi^{\left(+1\right)}\left(t\right)\right\rangle =\left\langle 0\right|_{\textrm{b}}\widehat{b}\left(\tau_{N}\right)\ldots\widehat{b}\left(\tau_{1}\right)\widehat{b}^{\dagger}\left|\varphi\left(t\right)\right\rangle \\
=M\left(\tau_{1}\right)\varphi_{N-1}\left(\tau_{2},\ldots,\tau_{N};t\right)+\sum_{1<p<N}M\left(\tau_{p}\right)\varphi_{N-1}\left(\tau_{1},\ldots,\tau_{p-1},\tau_{p+1},\ldots\tau_{N};t\right)\\
+M\left(\tau_{N}\right)\varphi_{N-1}\left(\tau_{1},\ldots,\tau_{N-1};t\right).
\end{multline}
Now, expanding the definition of $\widehat{H}_{\textrm{dress}}\left(z,t\right)$
in the equation (\ref{eq:joint_delay_time_eom_start}), and substituing
the derived here relations, one obtains the result (\ref{eq:delay_time_eq_1-1})-(\ref{eq:delay_time_eq_3-2}).

\section{\label{sec:REGULARIZATION-OF-THE}REGULARIZATION OF THE CHANNEL MEMORY
FUNCTION}

The channel decomposition of the waveguide memory function (\ref{eq:finite_band_memory_function})
is obtained by employing the representation of Bessel function 
\begin{equation}
J_{1}\left(z\right)=\frac{1}{2}H_{1}^{\left(1\right)}\left(z\right)+\frac{1}{2}H_{1}^{\left(2\right)}\left(z\right).\label{eq:bessel_decomposition}
\end{equation}
We obtain:
\begin{equation}
M\left(\tau\right)=e^{-i\left(\varepsilon-2h\right)\tau}M_{1}\left(\tau\right)+e^{-i\left(\varepsilon+2h\right)\tau}M_{2}\left(\tau\right),
\end{equation}
where
\begin{equation}
M_{1}\left(\tau\right)=e^{-i2h\tau}\frac{H_{1}^{\left(1\right)}\left(2h\tau\right)}{2h\tau},\label{eq:channel_1-1}
\end{equation}
\begin{equation}
M_{2}\left(\tau\right)=e^{+i2h\tau}\frac{H_{1}^{\left(2\right)}\left(2h\tau\right)}{2h\tau}.\label{eq:channel_2-1}
\end{equation}
The channel memory functions (\ref{eq:channel_1-1}) and (\ref{eq:channel_2-1})
have a singularity at zero delay times since
\begin{equation}
\frac{1}{z}H_{1}^{\left(1\right)}\left(z\right)\sim-\frac{2i}{\pi}\frac{1}{z^{2}}+\frac{i}{\pi}\ln(z)
\end{equation}
and
\begin{equation}
\frac{1}{z}H_{1}^{\left(2\right)}\left(z\right)\sim+\frac{2i}{\pi}\frac{1}{z^{2}}-\frac{i}{\pi}\ln\left(z\right).
\end{equation}
However, since the channel memory functions $M_{1}\left(\tau\right)$
and $M_{2}\left(\tau\right)$ are complex conjugated to each other,
they are defined up to an arbitrary real function $R\left(2h\tau\right)$,
Eqs. (\ref{eq:channel_1})-(\ref{eq:channel_2}). The choice of $R\left(2h\tau\right)$
is done from the requirement that $M_{k}\left(\tau\right)$ and its
first $n$ derivatives are finite at $\tau=0$. Here $n$ is not smaller
than the order of the required polynomial interpolation scheme (see
section \ref{subsec:The-numerical-coarsegraining}). In particular,
in this work we have made the choice
\begin{equation}
R\left(z\right)=\frac{1}{\pi}\left(\frac{2}{z^{2}}-\left(\ln\left(z\right)-\ln\left(2\right)+\gamma\right)\sum_{k=0}^{\infty}\alpha_{k}z^{2k}\right)e^{-8z^{2}}.
\end{equation}
Here 
\begin{equation}
\alpha_{0}=1,\,\,\,\alpha_{1}=\frac{252}{32},\,\,\,\alpha_{2}=\frac{35718}{1152},\,\,\,\alpha_{3}=\frac{17998824}{221184},\,\,\,\alpha_{4}=\frac{7085222460}{44236800},\ldots
\end{equation}
and $\gamma$ is the Euler--Mascheroni constant $\gamma=0.5772156\ldots$.

\section{\label{sec:SPLIT-OPERATOR-SIMULATION}SPLIT-OPERATOR SIMULATION}

First of all, observe that in order to complete the system of Eqs.
(\ref{eq:delay_time_channel_1})-(\ref{eq:delay_time_channel_4}),
we need to calculate self-consistently the shift field $\phi\left(t\right)$,
as the convolution
\begin{equation}
\phi\left(t\right)=-i\intop_{0}^{t}d\tau M\left(t-\tau\right)\overline{s}\left(\tau\right).\label{eq:shift-1}
\end{equation}
To do this in an efficient way, here we also apply the delay-time
approach. Namely, this convolution equation is equivalent to the following
delay-time equation:
\begin{equation}
\partial_{t}\phi\left(\tau;t\right)=\partial_{\tau}\phi\left(\tau;t\right)-iM\left(\tau\right)\overline{s}\left(\tau\right)\label{eq:self-consistent-shift-as-delay-time}
\end{equation}
for $\tau\geq0$ and with the initial condition $\phi\left(\tau;0\right)=0$.
Then, we identify: 
\begin{equation}
\phi\left(t\right)=\phi\left(0;t\right).
\end{equation}
In order to perform the coarsegraining of the memory tails, we rewrite
the equation (\ref{eq:self-consistent-shift-as-delay-time}) channel-wise:
\begin{equation}
\partial_{t}\phi^{\left(k\right)}\left(\tau;t\right)=\partial_{\tau}\phi^{\left(k\right)}\left(\tau;t\right)-iM_{k}\left(\tau\right)e^{+i\varepsilon_{k}t}\overline{s}\left(\tau\right).
\end{equation}
Then, the original shift field $\phi\left(t\right)$ is found as a
superposition of the channel shifts
\begin{equation}
\phi\left(t\right)=\sum_{k}e^{-i\varepsilon_{k}t}\varphi^{\left(k\right)}\left(0;t\right).
\end{equation}

All the unknowns in the system of differetial equations Eqs. (\ref{eq:delay_time_channel_1})-(\ref{eq:delay_time_channel_4})
can be collected into a single state vector 
\begin{equation}
X\left(t\right)=\left[\begin{array}{c}
\left|\varphi_{\textrm{0}}\left(t\right)\right\rangle \\
\left|\phi_{\textrm{1}}^{\left(k\right)}\left(\tau;t\right)\right\rangle \\
\left|\phi_{\textrm{2}}^{\left(kl\right)}\left(\tau_{1},\tau_{2};t\right)\right\rangle \\
\phi^{\left(k\right)}\left(\tau;t\right)
\end{array}\right].
\end{equation}
Then, the system of Eqs. (\ref{eq:delay_time_channel_1})-(\ref{eq:delay_time_channel_4})
can be compactly written as
\begin{equation}
\partial_{t}X\left(t\right)=F\left(t\right)\left[X\left(t\right)\right],
\end{equation}
where the right-hand side $F\left(t\right)$ is a time-dependent nonlinear
operator acting on $X\left(t\right)$. This operator $F$ is split
into a sum 
\begin{equation}
F\left(t\right)=S\left(t\right)+D\left(t\right)+H\left(t\right)+B\left(t\right),
\end{equation}
where each term is an operator. The operator $S$ represents the evolution
under the OQS Hamiltonian
\begin{equation}
\widehat{H}_{\textrm{stoch}}=\widehat{H}_{\textrm{s}}+\hat{s}\left(\xi\left(t\right)+\sum_{k}e^{+i\varepsilon_{k}t}\varphi^{\left(k\right)*}\left(0;t\right)\right),
\end{equation}
and the absorbtion from one-virtual quantum state: 
\begin{equation}
S\left(t\right)\left[\begin{array}{c}
\left|\varphi_{\textrm{0}}\left(t\right)\right\rangle \\
\left|\phi_{\textrm{1}}^{\left(k\right)}\left(\tau;t\right)\right\rangle \\
\left|\phi_{\textrm{2}}^{\left(kl\right)}\left(\tau_{1},\tau_{2};t\right)\right\rangle \\
\phi^{\left(k\right)}\left(\tau;t\right)
\end{array}\right]=-i\widehat{H}_{\textrm{stoch}}\left(t\right)\left[\begin{array}{c}
\left|\varphi_{\textrm{0}}\left(t\right)\right\rangle \\
\left|\phi_{\textrm{1}}^{\left(k\right)}\left(\tau;t\right)\right\rangle \\
\left|\phi_{\textrm{2}}^{\left(kl\right)}\left(\tau_{1},\tau_{2};t\right)\right\rangle \\
0
\end{array}\right]-i\left[\begin{array}{c}
\left(\widehat{s}^{\dagger}-\overline{s}^{*}\left(t\right)\right)\sum_{k}e^{-i\varepsilon_{k}t}\left|\phi_{1}^{\left(k\right)}\left(0;t\right)\right\rangle \\
0\\
0\\
0
\end{array}\right].
\end{equation}
The operator $D$ represents the free evolution of the bath and of
the self-consisent shift field:

\begin{equation}
D\left(t\right)\left[\begin{array}{c}
\left|\varphi_{\textrm{0}}\left(t\right)\right\rangle \\
\left|\phi_{\textrm{1}}^{\left(k\right)}\left(\tau;t\right)\right\rangle \\
\left|\phi_{\textrm{2}}^{\left(kl\right)}\left(\tau_{1},\tau_{2};t\right)\right\rangle \\
\phi^{\left(k\right)}\left(\tau;t\right)
\end{array}\right]=\left[\begin{array}{c}
0\\
\partial_{\tau}\left|\phi_{\textrm{1}}^{\left(k\right)}\left(\tau;t\right)\right\rangle \\
\left\{ \partial_{\tau_{1}}+\partial_{\tau_{2}}\right\} \left|\phi_{\textrm{2}}^{\left(kl\right)}\left(\tau_{1},\tau_{2};t\right)\right\rangle \\
\partial_{\tau}\phi^{\left(k\right)}\left(\tau;t\right)
\end{array}\right].
\end{equation}
The operator $H$ representes the emission of a virtual quantum and
of the self-consistent shift:
\begin{equation}
H\left(t\right)\left[\begin{array}{c}
\left|\varphi_{\textrm{0}}\left(t\right)\right\rangle \\
\left|\phi_{\textrm{1}}^{\left(k\right)}\left(\tau;t\right)\right\rangle \\
\left|\phi_{\textrm{2}}^{\left(kl\right)}\left(\tau_{1},\tau_{2};t\right)\right\rangle \\
\phi^{\left(k\right)}\left(\tau;t\right)
\end{array}\right]=\left[\begin{array}{c}
0\\
-iM_{k}\left(\tau\right)\widehat{s}e^{+i\varepsilon_{k}t}\left|\varphi_{\textrm{0}}\left(t\right)\right\rangle \\
-i\widehat{s}\frac{1}{1+\delta_{kl}}\left\{ e^{+i\varepsilon_{k}t}M_{k}\left(\tau_{1}\right)\left|\phi_{\textrm{1}}^{\left(l\right)}\left(\tau_{2};t\right)\right\rangle +e^{+i\varepsilon_{l}t}M_{l}\left(\tau_{2}\right)\left|\phi_{\textrm{1}}^{\left(k\right)}\left(\tau_{1};t\right)\right\rangle \right\} \\
-ie^{+i\varepsilon_{k}t}M_{k}\left(\tau\right)\overline{s}\left(t\right)
\end{array}\right].
\end{equation}
Finally, the operator $B$ represents the absorbtion of a virtual
quantum from higher-than-one virtual quantum states :
\begin{multline}
B\left(t\right)\left[\begin{array}{c}
\left|\varphi_{\textrm{0}}\left(t\right)\right\rangle \\
\left|\phi_{\textrm{1}}^{\left(k\right)}\left(\tau;t\right)\right\rangle \\
\left|\phi_{\textrm{2}}^{\left(kl\right)}\left(\tau_{1},\tau_{2};t\right)\right\rangle \\
\phi^{\left(k\right)}\left(\tau;t\right)
\end{array}\right]\\
=\left[\begin{array}{c}
0\\
-i\left(\widehat{s}^{\dagger}-s^{*}\left(t\right)\right)e^{-i\varepsilon_{k}t}\left|\phi_{2}^{\left(kk\right)}\left(0,\tau;t\right)+\phi_{2}^{\left(kk\right)}\left(\tau,0;t\right)\right\rangle -i\left(\widehat{s}^{\dagger}-s^{*}\left(t\right)\right)\sum_{l\neq k}e^{-i\varepsilon_{l}t}\left|\phi_{2}^{\left(kl\right)}\left(\tau,0;t\right)\right\rangle \\
0\\
0
\end{array}\right].
\end{multline}

Then, accroding to the split-operator method, up to the second order
in $dt$, the total evolution $\widehat{U}\left(dt;t\right)$ from
$t$ to $t+dt$ is represented as a symmetric composition
\begin{multline}
\widehat{U}\left(dt;t\right)=\exp\left(\frac{dt}{2}S\left(t+dt\right)\right)\circ\exp\left(\frac{dt}{2}H\left(t+dt\right)\right)\circ\exp\left(\frac{dt}{2}B\left(t+dt\right)\right)\circ\exp\left(dtD\left(t+\frac{dt}{2}\right)\right)\\
\circ\exp\left(\frac{dt}{2}B\left(t\right)\right)\circ\exp\left(\frac{dt}{2}H\left(t\right)\right)\circ\exp\left(\frac{dt}{2}S\left(t\right)\right),
\end{multline}
where the ``operator exponents'' have a symbolic meaning: by the
symbol 
\begin{equation}
\exp\left(dtA\left(\tau\right)\right)\left[Y\right]\label{eq:nonlinear_operator_exponent}
\end{equation}
we formally denote the solution $X\left(dt\right)$ of (nonlinear)
equation
\begin{equation}
\partial_{t}X\left(t\right)=A\left(\tau\right)\left[X\left(t\right)\right]\label{eq:nonlienar_operator_exponent_eq}
\end{equation}
with the initial condition 
\begin{equation}
X\left(0\right)=Y.
\end{equation}
Observe that in Eqs. (\ref{eq:nonlinear_operator_exponent}) and (\ref{eq:nonlienar_operator_exponent_eq}),
in the explicit time-dependence of $A$, the time argument is kept
fixed: $t=\tau$. Each such ``operator exponent'' is approximated
up to the second order in $dt$. The ``operator exponent'' $\exp\left(dtD\left(t+\frac{dt}{2}\right)\right)$
is implemented in a numerically exact way as a polynomial interpolation
in delay times.

\bibliographystyle{apsrev4-1}
\bibliography{references}

\end{document}